\documentclass[aps,prd,superscriptaddress,showpacs]{revtex4}
\usepackage{graphicx}
\usepackage{epstopdf}
\usepackage{amsmath}
\usepackage{amsfonts}
\usepackage{amssymb}
\usepackage{latexsym}

\begin{document}
\title{Remarks on nonlinear Electrodynamics}
\author{Patricio Gaete}\email{patricio.gaete@usm.cl}
 \affiliation{Departmento de F\'{i}sica and Centro Cient\'{i}fico-Tecnol\'ogico de Valpara\'{i}so, Universidad T\'{e}cnica Federico Santa Mar\'{i}a, Valpara\'{i}so, Chile}
\author{Jos\'{e} Helay\"{e}l-Neto}\email{helayel@cbpf.br}
\affiliation{Centro Brasileiro de Pesquisas F\'{i}sicas (CBPF), Rio de Janeiro, RJ, Brasil}

\begin{abstract}
We consider both generalized Born-Infeld and Exponential Electrodynamics. The field-energy of a point-like charge is finite only for Born-Infeld-like Electrodynamics. However, both Born-Infeld-type and Exponential Electrodynamics display the vacuum birefringence phenomenon. Subsequently, we calculate the lowest-order modifications to the interaction energy for both classes of Electrodynamics, within the framework of the gauge-invariant path-dependent variables formalism. These are shown to result in long-range ($1/r^5$- type) corrections to the Coulomb potential. Once again, for their non-commutative versions, the interaction energy is ultraviolet finite. 
\end{abstract}
\maketitle

\section{Introduction}

Amongst the most interesting of the phenomena predicted by Quantum Electrodynamics (QED) we may quote the photon-photon scattering in vacuum arising from the interaction of photons with virtual electron-positron pairs \cite{Breit-Wheeler,Adler,Costantini,Ruffini}. However, despite remarkable progresses \cite{Bamber,Burke,nphoton,Tommasini1,Tommasini2}, this prediction has not yet been confirmed. Nevertheless, this remarkable quantum characteristic of light remains a fascinating and challenging topic of research. In fact, it is conjectured that alternative scenarios such as Born-Infeld theory \cite{Born}, millicharged particles \cite{Gies} or axion-like particles \cite{Masso,Gaete1,Gaete2} may have more significant contributions to photon-photon scattering physics. 

It is worthy recalling, at this stage, that Born-Infeld (BI) Electrodynamics was proposed in 1934 in order to remove the singularities associated with charged point-like particles. Also, similarly to Maxwell Electrodynamics, Born-Infeld Electrodynamics displays no birefringence in vacuum. At the same time, BI Electrodynamics is distinguished, since BI-type effective actions arise in many different contexts in superstring theory \cite{Tseytlin,Pope}. Additionally, nonlinear electrodynamics (BI) have also been investigated in the context of gravitational physics \cite{Banerjee,Mann}. Actually, in addition to Born-Infeld theory, other types of nonlinear electrodynamics have been studied in the context of black hole physics \cite{Hendi,Zhao,Olea,Habib}. 

Meanwhile, recent experiments related to photon-photon interaction physics \cite{Bamber,Burke,nphoton,Tommasini1,Tommasini2}, have shown that the electrodynamics in vacuum is a nonlinear theory. Therefore, different models of nonlinear electrodynamics of vacuum deserve additional attention on the physical consequences presented by a particular nonlinear electrodynamics. In effect, our purpose here is to examine the properties of both Born-Infeld-like Electrodynamics and Exponential Electrodynamics.

On the other hand, we also recall that extensions of the Standard Model (SM) such as Lorentz invariance violating scenarios and fundamental length, have been object of intensive investigations over the past years \cite{AmelinoCamelia:2002wr,Jacobson:2002hd,Konopka:2002tt,Hossenfelder:2006cw,Nicolini:2008aj}. The main reason for this is that the SM does not include a quantum theory of gravitation, as well as the need to understand and to overcome difficulties theoretical in the quantum gravity research. An attempt along this direction has been to consider quantum field theories allowing non-commuting position operators  \cite{Witten:1985cc,Seiberg:1999vs,Douglas:2001ba,Szabo:2001kg,Gomis:2000sp,Bichl:2001nf}, where this non-commutativity is an intrinsic property of space-time. These studies were first made by using a start product (Moyal product). However, in recent years, a novel way to formulate noncommutative quantum field theory (or quantum field theory in the presence of a minimal length) has been initiated in \cite{Euro1,Euro2,Euro3}, which defines the fields as mean value over coherent states of the noncommutative plane. Later, it has been shown that the coherent state approach can be summarized through the introduction of a new multiplication rule which is known as Voros star-product \cite{Banerjee:2009xx}. Evidently, physics turns out be independent from the choice of the type of product \cite{Jabbari}. Subsequently, this new approach has been applied extensively to black holes physics \cite{Piero-Euro}.

With these considerations in mind, in a previous work \cite{Gaete-Helayel}, we have considered Logarithmic Electrodynamics, for which the field energy of a point-like charge is finite, as it happens in the case of the usual Born-Infeld electrodynamics. We have also shown that, contrary to the latter, Logarithmic  Electrodynamics displays the phenomenon of birefringence. Further, we have computed the lowest-order to the interaction energy for both Logarithmic Electrodynamics and for its non-commutative version, by using the gauge-invariant but path-dependent formalism. Our calculation has shown a long-range correction to the Coulomb potential for Logarithmic Electrodynamics. Interestingly enough, for its non-commutative version, the static potential becomes ultraviolet finite. From such a perspective, and given the outgoing experiments related to photon-photon interaction physics \cite{Bamber,Burke,nphoton,Tommasini1,Tommasini2}, the present work is an extension of our previous study \cite{Gaete-Helayel}. To do this, we shall work out the static potential for both Born-Infeld-like and Exponential Electrodynamics, using the gauge-invariant but path-dependent variables formalism, which is an alternative to the Wilson loop approach. 

Let us also mention here that Lagrangian densities of non-linear extensions of electrodynamics such as Born-Infeld-like Electrodynamics, whose Lagrangian density is built up with an arbitrary power of the electromagnetic invariants, have been considered in the context of black hole physics \cite{Hendi,Hassaine}. Whereas, in the context of single layer graphene, the effective action for the $(2+1)$ relativistic quantum electrodynamics is governed by a power $ {\raise0.5ex\hbox{$\scriptstyle 3$}
\kern-0.1em/\kern-0.15em\lower0.25ex\hbox{$\scriptstyle 4$}}$ \cite{Volovik,Katsnelson}. In addition, Exponential Electrodynamics has also been considered in physics of black holes \cite{Hendi}.

Our work is organized according to the following outline: in Section II, we consider Born-Infeld-like  Electrodynamics, show that it yields birefringence, compute the interaction energy for a fermion-antifermion pair and its version in the presence of a minimal length. In Section III, we repeat our analysis for Exponential Electrodynamics. Finally, in Section IV, we cast our Final Remarks.

\section{Born-Infeld-like model} 

As already stated, we now examine the interaction energy for Born-Infeld-like electrodynamics. To do this, we will calculate the expectation value of the energy operator $ H$ in the physical state $ |\Phi\rangle$, which we will denote by ${ \langle H\rangle_ \Phi}$. However, before going to the derivation of the interaction energy, we will describe briefly the model under consideration. The initial point of our analysis is the Lagrangian density:
\begin{equation}
{\cal L} = \beta ^2 \left\{ {1 - \left[ {1 + \frac{2}{{\beta ^2 }}{\cal F} - \frac{1}{{\beta ^4}}{\cal G}^2 } \right]^p } \right\}, \label{Exp05}
\end{equation}
where ${\cal F} = \frac{1}{4}F_{\mu \nu } F^{\mu \nu }$, $   
{\cal G} = \frac{1}{4}F_{\mu \nu } \tilde F^{\mu \nu }$. In the sequel, we shall justify why we confine ourselves to the domain $0 < p < 1$. As usually, $F_{\mu \nu }  = \partial _\mu  A_\nu   - \partial _\nu  A_\mu$ is the electromagnetic field strength tensor and $\tilde F^{\mu \nu }  = \frac{1}{2}\varepsilon ^{\mu \nu \rho \lambda } F_{\rho \lambda }$ is the dual electromagnetic field strength tensor. Before we proceed further, it is interesting to recall that Born and Infeld introduced their arbitrary Lagrangian, corresponding to $p={\raise0.5ex\hbox{$\scriptstyle 1$}
\kern-0.1em/\kern-0.15em\lower0.25ex\hbox{$\scriptstyle 2$}}$ in (\ref{Exp05}), in analogy to the relativistic Lagrangian of a free particle \cite{Dittrich}.

We start by observing that the field equations read:
\begin{equation}
\partial _\mu  \left[ {\frac{1}{\Gamma^{1-p} }\left( {F^{\mu \nu }  - \frac{1}{{\beta ^2 }}{\cal G}\tilde F^{\mu \nu } } \right)} \right] = 0, \label{Exp10}
\end{equation}
while the Bianchi identities are given by
\begin{equation}
\partial _\mu  \tilde F^{\mu \nu }  = 0, \label{Exp15}
\end{equation}
where
\begin{equation}
\Gamma  = 1 + \frac{{{2\cal F}}}{{\beta ^2 }} - \frac{{{\cal G}^2 }}{{\beta ^4}}. \label{Exp20}
\end{equation}

It follows from the above discussion that Gauss' law takes the form,
\begin{equation}
\nabla  \cdot {\bf D} = 0, \label{Exp20-5}
\end{equation}
where $\bf D$ is given by
\begin{equation}
{\bf D} = \frac{{{\bf E} + \frac{1}{{\gamma ^2 }}\left( {{\bf E} \cdot {\bf B}} \right){\bf B}}}{{[1 - \frac{{\left( {{\bf E}^2  - {\bf B}^2 } \right)}}{{\beta ^2 }} - \frac{1}{{\beta ^4}}\left( {{\bf E} \cdot {\bf B}} \right)^2]^{1-p} }}. \label{Exp20-10}
\end{equation}
From (\ref{Exp20-5}), for $J^0 (t,{\bf r}) = e\delta ^{\left( 3 \right)} \left( {\bf r} \right)$, it is clear that the ${\bf D}$-field is given by ${\bf D} = \frac{Q}{{r^2 }}\hat r$, where $Q = \frac{e}{{4\pi }}$. Considering the situation of a point-like charge, e, at the origin, the relation
\begin{equation}
\frac{Q}{{r^2 }} = \frac{{|{\bf E}|}}{{\left( {1 - \frac{{{\bf E}^2 }}{{\beta ^2 }}} \right)^{1 - p} }}, \label{Exp20-15}
\end{equation}
tells us that, for $r \to 0$, the electrostatic field is regular at the origin (where it acquires its maximum, $ 
E_{\max }  = \beta$) only with $p < 1$. On the other hand, $p < 0$ is excluded because there could exist field configurations for which the Lagrangian density would blow up. For $p > 1$, $E$ becomes singular at $r=0$; we then exclude this possibility, since we focus on solutions that are regular on the charge position. Our final choice is $0 < p < 1$, from the considerations above.

The physical picture is that for $0 < p < 1$, though the charged particle is point-like, its charge somehow spreads in a small region and it is screened by the polarization that results from the quantum effects which sum up to produce the effective Lagrangian (\ref{Exp05}).

From (\ref{Exp20-10}), we then easily verify that the electric field satisfy the equation
\begin{equation}
\frac{{D^\xi  }}{{\beta ^2 }}E^2  + E^\xi   - D^\xi   = 0,  \label{Exp25}
\end{equation}
where $\xi  = \frac{1}{{1 - p}}$. Now, it is worth noting that if $\xi$ is an integer, we obtain   
$p = {\raise0.5ex\hbox{$\scriptstyle 1$}
\kern-0.1em/\kern-0.15em
\lower0.25ex\hbox{$\scriptstyle 2$}},{\raise0.5ex\hbox{$\scriptstyle 2$}
\kern-0.1em/\kern-0.15em
\lower0.25ex\hbox{$\scriptstyle 3$}},{\raise0.5ex\hbox{$\scriptstyle 3$}
\kern-0.1em/\kern-0.15em
\lower0.25ex\hbox{$\scriptstyle 4$}},{\raise0.5ex\hbox{$\scriptstyle 4$}
\kern-0.1em/\kern-0.15em
\lower0.25ex\hbox{$\scriptstyle 5$}},....$. The $\xi=0$ and $\xi=1$ cases are excluded. We further note that negative values of $\xi$ are not permitted because it lead to $p > 1$. Evidently, when $\xi$ is an integer induces a richer dynamics than usual Born-Infeld electrodynamics $(p = {\raise0.5ex\hbox{$\scriptstyle 1$}\kern-0.1em/\kern-0.15em\lower0.25ex\hbox{$\scriptstyle 2$}})$. It is this aspect which we wish to investigate. In order to do so our considerations will be confined to the simplest possible case, namely,  $\xi=4$ (or $ p = {\raise0.5ex\hbox{$\scriptstyle 3$}\kern-0.1em/\kern-0.15em\lower0.25ex\hbox{$\scriptstyle 4$}}$). To accomplish this task, we begin by rewriting the Lagrangian density (\ref{Exp05}) as 
\begin{equation}
{\cal L} = \beta ^2 \left\{ {1 - \left[ {1 + \frac{1}{{3\beta ^2 }}F_{\mu \nu }^2  - \frac{1}{{24\beta ^4}}\left( {F_{\mu \nu } \tilde F^{\mu \nu } } \right)^2 } \right]^{{\raise0.5ex\hbox{$\scriptstyle 3$}
\kern-0.1em/\kern-0.15em
\lower0.25ex\hbox{$\scriptstyle 4$}}} } \right\}.  \label{Exp30}
\end{equation}

This then implies that the electric field reduces to
\begin{equation}
{\bf E} = \beta \sqrt 3 Q\frac{1}{{\sqrt {Q^2  + \sqrt {Q^4  + 9\beta ^4 r^8 } } }}\hat r, \label{Exp35}
\end{equation}
From this expression, it should be clear that the electric field of a point-like particle is maximum at the origin, $E_{\max }  = \beta \sqrt {\frac{3}{2}}$; in the usual Born-Infeld Electrodynamics, ${E_{\max }} = \beta$. Notice that $E_{\max }$ is no longer $\beta$ ($E_{\max}  = \beta \sqrt {\frac{3}{2}}$) simply because we have changed our coefficients in (\ref{Exp25}) (they are no longer the ones defined in (\ref{Exp05})).

In order to properly discuss the propagation of electromagnetic waves within this context, it is advantageous to introduce the vectors ${\bf D} = {{\partial {\cal L}} \mathord{\left/
 {\vphantom {{\partial L} {\partial {\bf E}}}} \right.
 \kern-\nulldelimiterspace} {\partial {\bf E}}}$ and ${\bf H} =  - {{\partial {\cal L}} \mathord{\left/
 {\vphantom {{\partial L} {\partial {\bf B}}}} \right.
 \kern-\nulldelimiterspace} {\partial {\bf B}}}$:
\begin{equation}
{\bf D} =
\frac{1}{{\Gamma ^{{\raise0.5ex\hbox{$\scriptstyle 1$}
\kern-0.1em/\kern-0.15em
\lower0.25ex\hbox{$\scriptstyle 4$}}} }}\left( {{\bf E} + \frac{{{\bf B}\left( {{\bf E} \cdot {\bf B}} \right)}}{{\beta ^2 }}} \right), \label{Exp40}
\end{equation}
and
\begin{equation}
{\bf H} = 
\frac{1}{{\Gamma ^{{\raise0.5ex\hbox{$\scriptstyle 1$}
\kern-0.1em/\kern-0.15em
\lower0.25ex\hbox{$\scriptstyle 4$}}} }}
\left( {{\bf B} - \frac{{{\bf E}\left( {{\bf E} \cdot {\bf B}} \right)}}{{\beta ^2 }}} \right), \label{Exp45}
\end{equation}
where $ 
\Gamma  = 1 + \frac{2}{{3\beta ^2 }}\left( {{\bf B}^2  - {\bf E}^2 } \right) - \frac{2}{{3\beta ^4 }}\left( {{\bf E} \cdot {\bf B}} \right)^2$. This means that the corresponding equations of motion are written as
\begin{equation}
\nabla  \cdot {\bf D} = 0, \  \  \
\frac{{\partial {\bf D}}}{{\partial t}} - \nabla  \times {\bf H} = 0, \label{Exp50}
\end{equation}
and
\begin{equation}
\nabla  \cdot {\bf B} = 0, \  \  \
\frac{{\partial {\bf B}}}{{\partial t}} + \nabla  \times {\bf E} = 0. \label{Exp55}
\end{equation}

From  Eqs. (\ref{Exp40}) and (\ref{Exp45}), it is clear that we can also write the electric permitivity 
$\varepsilon _{ij}$ and the inverse magnetic permeability $\left( {\mu}^{-1}  \right)_{ij}$ tensors of the vacuum. In effect,
\begin{equation}
\varepsilon _{ij}  = 
\frac{1}{{\Gamma ^{{\raise0.5ex\hbox{$\scriptstyle 1$}
\kern-0.1em/\kern-0.15em
\lower0.25ex\hbox{$\scriptstyle 4$}}} }}
\left( {\delta _{ij}  + \frac{1}{{\beta ^2 }}B_i B_j } \right), \ \ \
\left( {\mu}^{-1}  \right)_{ij}  = 
\frac{1}{{\Gamma ^{{\raise0.5ex\hbox{$\scriptstyle 1$}
\kern-0.1em/\kern-0.15em
\lower0.25ex\hbox{$\scriptstyle 4$}}} }}\left( {\delta _{ij}  - \frac{1}{{\beta ^2 }}E_i E_j } \right), \label{Exp60}
\end{equation}
with $D_i  = \varepsilon _{ij} E_j$ and $B_i  = \mu _{ij} H_j $.

Following our earlier procedure \cite{Gaete-Helayel}, we must linearize the above equations. To this end one considers a weak electromagnetic wave $({\bf E_p}, {\bf B_p})$ propagating in the presence of a strong constant external field $({\bf E_0}, {\bf B_0})$. We see, therefore, that for the case of a purely magnetic field (${\bf E_0}=0$), the vectors ${\bf D}$ and ${\bf H}$ reduce to   
\begin{equation}
{\bf D} = \frac{1}{{\left( {1 + {\textstyle{2 \over 3}}{\textstyle{{{\bf B}_0^2 } \over {\beta ^2 }}}} \right)^{{\raise0.5ex\hbox{$\scriptstyle 1$}
\kern-0.1em/\kern-0.15em
\lower0.25ex\hbox{$\scriptstyle 4$}}} }}\left[ {{\bf E}_p  + \frac{1}{{\beta ^2 }}\left( {{\bf E}_p  \cdot {\bf B}_0 } \right){\bf B}_0 } \right], \label{Exp65}
\end{equation}
and
\begin{equation}
{\bf H} = \frac{1}{{\left( {1 + {\textstyle{2 \over 3}}{\textstyle{{{\bf B}_0^2 } \over {\beta ^2 }}}} \right)^{{\raise0.5ex\hbox{$\scriptstyle 1$}
\kern-0.1em/\kern-0.15em
\lower0.25ex\hbox{$\scriptstyle 4$}}} }}\left[ {{\bf B}_p  - \frac{1}{{3\beta ^2 \left( {1 + {\textstyle{2 \over 3}}{\textstyle{{{\bf B}_0^2 } \over {\beta ^2 }}}} \right)}}\left( {{\bf B}_p  \cdot {\bf B}_0 } \right){\bf B}_0 } \right], \label{Exp70} 
\end{equation}
where we have keep only linear terms in ${\bf E_p}$, ${\bf B_p}$. Now, by considering the $z$ axis as the direction of the external magnetic field (${\bf B_0}  = B_0 {\bf e}_3$) and assuming that the light wave moves along the $x$ axis, the decomposition into a plane wave for the fields ${\bf E}_p$ and ${\bf B}_p$ can be written as
\begin{equation}
{{\bf E_p}}\left( {{\bf x}
,t} \right) = {\bf E}
{e^{ - i\left( {wt - {\bf k} \cdot {\bf x}} \right)}}, \ \ \
{{\bf B_p}}\left( {{\bf x},t} \right) = {\bf B}{e^{ - i\left( {wt - {\bf k} \cdot {\bf x}} \right)}}. \label{Exp75}
\end{equation}
Once this is done, we arrive at the following Maxwell equations: 
\begin{equation}
\left( {\frac{{{k^2}}}{{{w^2}}} - {\varepsilon _{22}}{\mu _{33}}} \right){E_2} = 0, \label{Exp80}
\end{equation}  
and
\begin{equation}
\left( {\frac{{{k^2}}}{{{w^2}}} - {\varepsilon _{33}}{\mu _{22}}} \right){E_3} = 0. \label{Exp85}
\end{equation}

Thus, from a physical point of view, we have two different situations: First, if ${\bf E}\ \bot \ {\bf B}_0$ (perpendicular polarization), from (\ref{Exp85}) $E_3=0$, and from (\ref{Exp80}) we get $\frac{{{k^2}}}{{{w^2}}} = {\varepsilon _{22}}{\mu _{33}}$. As a result, the dispersion relation of the photon may be written as 
\begin{equation}
n_ \bot   = \sqrt {\frac{{1 + \frac{{2{\bf B}_0^2 }}{{3\beta ^2 }}}}{{1 + \frac{{{\bf B}_0^2 }}{{3\beta ^2 }}}}} . \label{Exp90}
\end{equation}
Second, if ${\bf E}\ || \ {\bf B}_0$ (parallel polarization), from (\ref{Exp80}) $E_2=0$, and from (\ref{Exp85}) we get $\frac{{{k^2}}}{{{w^2}}} = {\varepsilon _{33}}{\mu _{22}}$. In this case, the corresponding dispersion relation becomes
\begin{equation}
n_\parallel   = \sqrt {1 + \frac{{{\bf B}_0^2 }}{{\beta ^2 }}}.  \label{Exp95}
\end{equation}
Thus we learn that the vacuum birefringence phenomenon is present, or what is the same, electromagnetic waves with different polarizations have different velocities.
Incidentally, it is of interest to notice that in Born-Infeld theory, which is described by a square root instead a power (${\raise0.5ex\hbox{$\scriptstyle 3$}\kern-0.1em/\kern-0.15em
\lower0.25ex\hbox{$\scriptstyle 4$}}$) as in (\ref{Exp30}), the phenomenon of birefringence is absent. However,  in the case of a generalized Born-Infeld electrodynamics \cite{Kruglov2}, which contains two different parameters, again the phenomenon of birefringence is present.

We now turn to the problem of obtaining the interaction energy between static point-like sources for a Born-Infeld-like model, our analysis follows closely that of references \cite{Gaete:2011ka,Gaete:2012yu}.
As already expressed, the corresponding theory is governed by the Lagrangian density (\ref{Exp30}), that is,
\begin{equation}
{\cal L} = \beta ^2 \left\{ {1 - \left[ {1 + \frac{1}{{3\beta ^2 }}F_{\mu \nu }^2  - \frac{1}{{24\beta ^2 \gamma ^2 }}\left( {F_{\mu \nu } \tilde F^{\mu \nu } } \right)^2 } \right]^{{\raise0.5ex\hbox{$\scriptstyle 3$}
\kern-0.1em/\kern-0.15em
\lower0.25ex\hbox{$\scriptstyle 4$}}} } \right\}.  \label{Exp100}
\end{equation}

As in our previous works \cite{Gaete-Helayel}, in order to handle the exponent $  
{\raise0.5ex\hbox{$\scriptstyle 3$}
\kern-0.1em/\kern-0.15em
\lower0.25ex\hbox{$\scriptstyle 4$}}$ in (\ref{Exp100}), we incorporate an auxiliary field $v$ such that its equation of motion gives back the original theory. This allows us to write the Lagrangian density as
\begin{equation}
{\cal L} = \beta ^2  - 3\beta ^2 v - vF_{\mu \nu }^2  + \frac{v}{{8\gamma ^2 }}\left( {F_{\mu \nu } \tilde F^{\mu \nu } } \right)^2  - \frac{{\beta ^2 }}{{4^4 }}\frac{1}{{v^3 }}. \label{Exp105}
\end{equation}

We now carry out a Hamiltonian analysis of this theory. The canonical momenta are found to be $    
\Pi ^\mu   =  - 4v\left( {F^{0\mu }  - \frac{1}{{4\gamma ^2 }}F_{\alpha \beta } \tilde F^{\alpha \beta } \tilde F^{0\mu } } \right)$, and one immediately identifies the two primary constraints 
$\Pi ^0  = 0$ and $p \equiv \frac{{\partial L}}{{\partial \dot v}} = 0$. The canonical Hamiltonian corresponding of the model can be worked out as usual and is given by the expression:
\begin{equation}
H_C  = \int {d^3 x} \left\{ {\Pi _i \partial ^i A^0  + \frac{1}{{8v}}{\bf \Pi} ^2  - \beta ^2  + 3\beta ^2 v + 2v{\bf B}^2  + \frac{{\beta ^2 }}{{4^4 }}\frac{1}{{v^3 }} - \frac{{\left( {{\bf \Pi}  \cdot {\bf B}} \right)^2 }}{{8v\gamma ^2 \left( {1 + \frac{{{\bf B}^2 }}{{\gamma ^2 }}} \right)}}} \right\}. \label{Exp110}
\end{equation}

Requiring the primary constraint $\Pi^{0}$ to be preserved in time, one obtains the secondary constraint $\Gamma _1  = \partial _i \Pi ^i  = 0$. Similarly for the constraint $p$, we get the auxiliary field $v$ as 
\begin{eqnarray}
v&=&\sqrt {\frac{1}{{16\gamma ^2 \det D\left( {2{\bf B}^2  + 3\beta ^2 } \right)}}} \nonumber\\
&\times&
\sqrt {\left[ {\left( {{\bf \Pi} ^2 \gamma ^2 \det D - \left( {{\bf \Pi}  \cdot {\bf B}} \right)^2 } \right) + \sqrt {\left( {{\bf \Pi} ^2 \gamma ^2 \det D - \left( {{\bf \Pi}  \cdot {\bf B}} \right)^2 } \right)^2  + 3\beta ^2 \left( {2{\bf B}^2  + 3\beta ^2 } \right)\left( {\gamma ^2 \det D} \right)^2 } } \right]}, \label{Exp115}
\end{eqnarray}
which will be used to eliminate $v$. Next, the corresponding total (first-class) Hamiltonian that generates the time evolution of the dynamical variables is $  
H = H_C  + \int {d^3 x} \left( {u_0(x) \Pi_0(x)  + u_1(x) \Gamma _1(x) } \right)$, where $u_o(x)$ and $u_1(x)$ are the Lagrange multiplier utilized to implement the constraints. It should be noted that $\dot A_0 \left( x \right) = \left[ {A_0 \left( x \right),H} \right] = u_0 \left( x \right)$, which is an arbitrary function. Since $\Pi^0=0$ always, neither $A^0$ nor $\Pi^0$ are of interest in describing the system and may be discarded from the theory. Thus the Hamiltonian is now given as
\begin{equation}
H= \int {d^3 x} \left\{ {w(x)\Pi _i \partial ^i + \frac{1}{{8v}}{\bf \Pi} ^2  - \beta ^2  + 3\beta ^2 v + 2v{\bf B}^2  + \frac{{\beta ^2 }}{{4^4 }}\frac{1}{{v^3 }} - \frac{{\left( {{\bf \Pi}  \cdot {\bf B}} \right)^2 }}{{8v\gamma ^2 \left( {1 + \frac{{{\bf B}^2 }}{{\gamma ^2 }}} \right)}}} \right\}. \label{Exp120}
\end{equation}
where $w(x) = u_1 (x) - A_0 (x)$ and $v$ is given by (\ref{Exp115}).

At this stage, in accordance with the Hamiltonian analysis, we must fix the gauge. A particularly
convenient gauge-fixing condition is
\begin{equation}
\Gamma _2 \left( x \right) \equiv \int\limits_{C_{\xi x} } {dz^\nu
} A_\nu \left( z \right) \equiv \int\limits_0^1 {d\lambda x^i }
A_i \left( {\lambda x} \right) = 0. \label{Exp125}
\end{equation}
where  $\lambda$ $(0\leq \lambda\leq1)$ is the parameter describing
the spacelike straight path $ x^i = \xi ^i  + \lambda \left( {x -
\xi } \right)^i $, and $ \xi $ is a fixed point (reference point).
There is no essential loss of generality if we restrict our
considerations to $ \xi ^i=0 $. With this, we arrive at the only non-vanishing equal-time Dirac bracket for the canonical variables
\begin{equation}
\left\{ {A_i \left( x \right),\Pi ^j \left( y \right)} \right\}^ *
= \delta _i^j \delta ^{\left( 3 \right)} \left( {x - y} \right) -
\partial _i^x \int\limits_0^1 {d\lambda x^i } \delta ^{\left( 3
\right)} \left( {\lambda x - y} \right). \label{Exp130}
\end{equation}

Having thus outlined the necessary aspects of quantization, we now proceed to compute the interaction energy for the model under consideration. Recalling again that we will work out the expectation value of the energy operator $H$ in the physical state $\left| \Phi  \right\rangle$, where the physical states $\left| \Phi  \right\rangle$ are gauge-invariant ones. We also recall that the stringy gauge-invariant state is given by:
\begin{equation}
\Pi _i \left( x \right)\left| {\overline \Psi \left( \mathbf{y }\right)\Psi
\left( {\mathbf{y}^ \prime } \right)} \right\rangle = \overline \Psi \left( 
\mathbf{y }\right)\Psi \left( {\mathbf{y}^ \prime } \right)\Pi _i \left( x
\right)\left| 0 \right\rangle + e\int_ {\mathbf{y}}^{\mathbf{y}^ \prime } {\
dz_i \delta ^{\left( 3 \right)} \left( \mathbf{z - x} \right)} \left| \Phi
\right\rangle.  \label{Exp135}
\end{equation}
This then implies that the lowest-order corrections to the modification interaction energy may be written as
\begin{equation}
\left\langle H \right\rangle _\Phi   = \left\langle H \right\rangle _0  + \left\langle H \right\rangle _\Phi ^{\left( 1 \right)}, \label{Exp140}
\end{equation}
where $\left\langle H \right\rangle _0  = \left\langle 0 \right|H\left| 0 \right\rangle$. The $\left\langle H \right\rangle _\Phi ^{\left( 1 \right)}$-term is given by
\begin{equation}
\left\langle H \right\rangle _\Phi ^{\left( 1 \right)}  = \left\langle \Phi  \right|\int {d^3 x} \left\{ {\frac{1}{2}{\bf \Pi} ^2  - \frac{1}{{12\beta ^2 }}{\bf \Pi} ^4 } \right\}\left| \Phi  \right\rangle. \label{Exp145} 
\end{equation}
It is, up to a numerical factor, just the interaction energy for the standard Born-Infeld theory. In view of this situation, the static potential turns out to be
\begin{equation}
V =  - \frac{{e^2 }}{{4\pi }}\frac{1}{L}\left( {1 - \frac{{e^2 }}{{480\pi ^2 \beta ^2 }}\frac{1}{{L^4 }}} \right).
\label{Exp150}
\end{equation}

As a second derivation of our previous result, it may be recalled that \cite{PG1,PG2}
\begin{equation}
V \equiv e\left( {{\cal A}_0 \left( {\bf 0} \right) - {\cal A}_0 \left( {\bf L} \right)} \right), \label{Exp150}
\end{equation}
where the physical scalar potential is given by
\begin{equation}
{\cal A}_0 (t,{\bf r}) = \int_0^1 {d\lambda } r^i E_i (t,\lambda
{\bf r}). \label{Exp155}
\end{equation}
This equation follows from the vector gauge-invariant field expression
\begin{equation}
{\cal A}_\mu  (x) \equiv A_\mu  \left( x \right) + \partial _\mu  \left( { - \int_\xi ^x {dz^\mu  A_\mu  \left( z \right)} } \right), \label{Exp160}
\end{equation}
where the line integral is along a spacelike path from the point $\xi$ to $x$, on a fixed slice time. It should be noted that the gauge-invariant variables (\ref{Exp160}) commute with the sole first constraint (GaussÕ law), showing in this way that these fields are physical variables. In passing we note that
GaussÕs law for the present theory $\partial _i \Pi ^i  = J^0$, where we have included the external current $J^0$ to represent the presence of two opposite charges. Now, using the previous result  (\ref{Exp35}), we note that the scalar potential for Born-Infeld-like electrodynamics, at leading order in $\beta$, is expressed as 
\begin{equation}
{\cal A}_0 \left( {t,{\bf r}} \right) =  - \frac{e}{{4\pi r}}\int_0^1 {d\lambda } \left\{ {\frac{1}{{\lambda ^2 }} - \frac{e^2}{96 \beta^2\pi^2r^4}\frac{{1 }}{{\lambda ^6 }}} \right\}, \label{Exp165}
\end{equation}

In this way, by employing Eq. (\ref{Exp150}), the potential for a pair of static point-like opposite charges located at $\bf 0$ and $\bf L$, is given by
\begin{equation}
V =  - \frac{{e^2 }}{{4\pi
}}\frac{1}{L}\left( {1 - \frac{{e^2 }}{{480\pi ^2 \beta ^2
}}\frac{1}{{L^4 }}} \right). \label{Exp170}
\end{equation}
One immediately sees that this is exactly the profile obtained for Born-Infeld electrodynamics. The point we wish to emphasize, however, is that Born-Infeld-like electrodynamics also has a rich structure reflected by its long-range correction to the Coulomb potential. 

Again, as in the case of both Born-Infeld and logarithmic electrodynamics \cite{Pato,Gaete-Helayel} have a finite electric field at the origin, we find that the interaction energy between two test charges at leading order in $\beta$ is not finite at the origin. Following our earlier line of argument \cite{Gaete:2011ka,Gaete:2012yu}, we shall give a concise description of Born-Infeld-like electrodynamics defined in a non-commutative geometry. In such a case, Gauss' law reads
\begin{equation}
\partial _i \Pi ^i  = e \ e^{ - \theta \nabla ^2 } \delta ^{\left( 3 \right)} \left( {\bf x} \right), \label{Exp175}
\end{equation}
thus $
\Pi ^i  = -\frac{{2e}}{{\sqrt \pi  }}\frac{{\hat r^i }}{{r^2 }}\gamma \left( {{\raise0.5ex\hbox{$\scriptstyle 3$}
\kern-0.1em/\kern-0.15em
\lower0.25ex\hbox{$\scriptstyle 2$}},{\raise0.5ex\hbox{$\scriptstyle {r^2 }$}
\kern-0.1em/\kern-0.15em
\lower0.25ex\hbox{$\scriptstyle {4\theta }$}}} \right)$,
with $r = |{\bf r}|$. While $\gamma \left( {{\raise0.5ex\hbox{$\scriptstyle 3$}
\kern-0.1em/\kern-0.15em
\lower0.25ex\hbox{$\scriptstyle 2$}},{\raise0.5ex\hbox{$\scriptstyle {r^2 }$}
\kern-0.1em/\kern-0.15em
\lower0.25ex\hbox{$\scriptstyle {4\theta }$}}} \right)$ is the lower incomplete Gamma function defined by 
$\gamma \left( {{\raise0.5ex\hbox{$\scriptstyle a$}
\kern-0.1em/\kern-0.15em
\lower0.25ex\hbox{$\scriptstyle b$}},x} \right) \equiv \int_0^x {\frac{{du}}{u}} u^{{\raise0.5ex\hbox{$\scriptstyle a$}
\kern-0.1em/\kern-0.15em
\lower0.25ex\hbox{$\scriptstyle b$}}} e^{ - u}$. 

By proceeding in the same way as in \cite{Gaete-Helayel}, we obtain the static potential for two opposite 
charges $e$ located at $\bf 0$ and $\bf L$:
\begin{equation}
V =  - \frac{{e^2 }}{{4\pi ^{{\raise0.5ex\hbox{$\scriptstyle 3$}
\kern-0.1em/\kern-0.15em
\lower0.25ex\hbox{$\scriptstyle 2$}}} }}\frac{1}{L}\left[ {\gamma \left( {{\raise0.5ex\hbox{$\scriptstyle 1$}
\kern-0.1em/\kern-0.15em
\lower0.25ex\hbox{$\scriptstyle 2$}},{\raise0.5ex\hbox{$\scriptstyle {L^2 }$}
\kern-0.1em/\kern-0.15em
\lower0.25ex\hbox{$\scriptstyle {4\theta }$}}} \right) + \frac{{16e^2 }}{{3\beta ^2 }}L\ \hat r^i \int_{\bf 0}^{\bf L} {dy^i } \frac{1}{{y^6 }}\gamma ^3 \left( {{\raise0.5ex\hbox{$\scriptstyle 3$}
\kern-0.1em/\kern-0.15em
\lower0.25ex\hbox{$\scriptstyle 2$}},{\raise0.5ex\hbox{$\scriptstyle {y^2 }$}
\kern-0.1em/\kern-0.15em
\lower0.25ex\hbox{$\scriptstyle {4\theta }$}}} \right)} \right], \label{LnE165}
\end{equation}
which is finite for $L \to 0$. Again, in the limit  $\theta \to 0$, we recover our above result. 

We would like to point out that we are not here going to calculate the electrostatic energy stored in a region corresponding to the Compton wavelength of the electron, $m_e^{ - 1}$, because we know that the electron mass is not originated from its electrostatic field; it rather comes from the Yukawa coupling between the electron and the Higgs fields and the spontaneous breaking down of the $  
SU_L \left( 2 \right) \times U_Y \left( 1 \right)$-symmetry to the electromagnetic $U(1)$; actually, $
m_e  = y_e \left\langle H \right\rangle$, where $y_e$ is the electron's Yukawa coupling and $
\left\langle H \right\rangle$ the order parameter of the breaking of electroweak symmetry.

\section{Exponential electrodynamics}

Our next undertaking is to use the ideas of the previous section in
order to consider exponential electrodynamics. In such a case the Lagrangian density reads \cite{Hendi}:
\begin{equation}
{\cal L} = \beta ^2 (e^{ - {\raise0.7ex\hbox{$X$} \!\mathord{\left/
 {\vphantom {X {\beta ^2 }}}\right.\kern-\nulldelimiterspace}
\!\lower0.7ex\hbox{${\beta ^2 }$}}}  - 1), \label{Expo05}
\end{equation}
where
\begin{equation}
X = {\cal F} - \frac{{{\cal G}^2 }}{{2\beta ^2 }}. \label{Expo10}
\end{equation}

As in the previous section, before we proceed to work out explicitly the interaction energy, we shall begin by considering the equations of motion. Thus we have
\begin{equation}
\partial _\mu  \left[ {\left( {F^{\mu \nu }  - \frac{1}{{\beta ^2 }}{\cal G}\tilde F^{\mu \nu } } \right)e^{ - {\raise0.7ex\hbox{$X$} \!\mathord{\left/
 {\vphantom {X {\beta ^2 }}}\right.\kern-\nulldelimiterspace}
\!\lower0.7ex\hbox{${\beta ^2 }$}}} } \right] = 0,  \label{Expo15}
\end{equation}
while the Bianchi identities are
\begin{equation}
\partial _\mu  \tilde F^{\mu \nu }  = 0. \label{Expo20}
\end{equation}

As before, we also find that  
\begin{equation}
\nabla  \cdot {\bf D} = 0, \label{Expo25} 
\end{equation}
where $\bf D$ takes the form
\begin{equation}
{\bf D} = e^{ - {\raise0.7ex\hbox{$X$} \!\mathord{\left/
 {\vphantom {X {\beta ^2 }}}\right.\kern-\nulldelimiterspace}
\!\lower0.7ex\hbox{${\beta ^2 }$}}} \left( {{\bf E} + \frac{1}{{\beta ^2 }}\left( {{\bf E} \cdot {\bf B}} \right){\bf B}} \right), \label{Expo30}
\end{equation}
where $X =  - \frac{1}{2}\left( {{\bf E}^2  - {\bf B}^2 } \right) - \frac{1}{{2\beta ^2 }}\left( {{\bf E} \cdot {\bf B}} \right)^2$.
Again, for $J^0 (t,{\bf r}) = e\delta ^{\left( 3 \right)} \left( {\bf r} \right)$, the ${\bf D}$-field is given by ${\bf D} = \frac{Q}{{r^2 }}\hat r$, where $Q = \frac{e}{{4\pi }}$. Moreover, from (\ref{Expo30}) we observe that the electric field satisfy the equation
\begin{equation}
{\bf D} = {\bf E}e^{{\raise0.7ex\hbox{${{\bf E}^2 }$} \!\mathord{\left/
 {\vphantom {{E^2 } {2\beta ^2 }}}\right.\kern-\nulldelimiterspace}
\!\lower0.7ex\hbox{${2\beta ^2 }$}}}, \label{Expo35}
\end{equation}
from which follows that 
\begin{equation}
{\bf E} = \beta \sqrt {W\left( {\frac{{Q^2 }}{{\beta ^2 r^4 }}} \right)}\hat r, \label{Expo40}
\end{equation}
where ${W\left( {\frac{{Q^2 }}{{\beta ^2 r^4 }}} \right)}$ is the Lambert function or Product Log function. In the limit $\beta  \to \infty$, the electric field reduces to
\begin{equation}
{\bf E} = \frac{Q}{{r^2 }}\sqrt {1 + \frac{{Q^2 }}{{\beta ^2 r^4 }}}\hat r. \label{Expo45}
\end{equation}
It is interesting to note that, for exponential electrodynamics, the electric field is not finite at the origin.

The physical understanding for a non-regular solution for $r=0$ is that Exponential Electrodynamics is actually a power series expansion in $\cal F$ and ${\cal G}^2$ and, then, as we have discussed in Section II, positive powers in $\cal F$ and $\cal G$ do not yet a finite field on the charge's position. We should point out that the Born-Infeld case ($p = {\raise0.5ex\hbox{$\scriptstyle 1$}\kern-0.1em/\kern-0.15em
\lower0.25ex\hbox{$\scriptstyle 2$}}$) comes out from the vacuum polarization as a quantum effect of virtual pair production and annihilation, which is responsible for the screening of a charge in a polarized vacuum. For $0 < p < 1$, the regime of screening is still valid. However, $p > 1$ is outside this regime and this is why the case of the Exponential Electrodynamics does not exhibit a regular electrostatic field at the charge's position. We point out, in this context, the work of Ref. \cite{Costa}, where a quartic model in $F_{\mu\nu}$ is considered and, though the field is not regular for $r=0$, the finiteness of the field energy is ensured.

Let us next consider the propagation of electromagnetic waves in exponential electrodynamics. Once again following the same steps leading to (\ref{Exp90}) and (\ref{Exp95}), in the present case, the dispersion relations read
\begin{equation}
n_ \bot   = \sqrt {\frac{{1 - \frac{{{\bf B}_0^2 }}{{2\beta ^2 }}}}{{1 - \frac{3{{\bf B}_0^2 }}{{2\beta ^2 }}}}} ,\label{Expo46}
\end{equation}
and
\begin{equation}
n_\parallel   = \sqrt {1 + \frac{{{\bf B}_0^2 }}{{\beta ^2 }}}. \label{Expo47}
\end{equation}
With this then, we see that the vacuum birefringence phenomenon is present. 

We now turn our attention to the calculation of the interaction energy between static point-like sources for exponential electrodynamics (\ref{Expo05}). Now, proceeding as before, we introduce the auxiliary field $v$, in order to handle the exponential in (\ref{Expo05}). In so doing, we get 
\begin{equation}
{\cal L} =  - \beta ^2 v\ln v + \beta ^2 v - \beta ^2  - \frac{v}{4}F_{\mu \nu }^2  + \frac{v}{{32\beta ^2 }}\left( {F_{\mu \nu } \tilde F^{\mu \nu } } \right)^2. \label{Expo50}
\end{equation}

It is once again straightforward to apply the gauge-invariant formalism discussed in the preceding section. For this purpose, we shall first carry out its Hamiltonian analysis. The canonical momenta read $    
\Pi ^\mu   =  - v\left( {F^{0\mu }  - \frac{1}{{4\beta ^2 }}F_{\alpha \beta } \tilde F^{\alpha \beta } \tilde F^{0\mu } } \right)$, and one immediately identifies the two primary constraints 
$\Pi ^0  = 0$ and $p \equiv \frac{{\partial L}}{{\partial \dot v}} = 0$. The canonical Hamiltonian corresponding to (\ref{Expo05}) is 
\begin{equation}
H_C  = \int {d^3 x} \left\{ {\Pi _i \partial ^i A_0  + \frac{1}{{2v}}{\bf \Pi} ^2  + \frac{v}{2}{\bf B}^2  - \frac{{\left( {{\bf B} \cdot {\bf \Pi} } \right)^2 }}{{2\beta ^2 v\det D}} + \beta ^2  + \beta ^2 v\ln v - \beta ^2 v} \right\}. \label{Expo55}
\end{equation}

Requiring the primary constraint $\Pi^{0}$ to be preserved in time, one obtains the secondary constraint $\Gamma _1  = \partial _i \Pi ^i  = 0$. Similarly for the constraint $p$, we get the auxiliary field $v$ as 
\begin{equation}
v = \frac{{\sqrt {\frac{{{\bf \Pi} ^2 }}{{\beta ^2 }} - \frac{{\left( {{\bf B} \cdot {\bf \Pi} } \right)^2 }}{{\beta ^4 \det D}}} }}{{\sqrt {W\left( {\left[ {\frac{{{\bf \Pi} ^2 }}{{\beta ^2 }} - \frac{{\left( {{\bf B} \cdot {\bf \Pi} } \right)^2 }}{{\beta ^4 \det D}}} \right]e^{{\raise0.7ex\hbox{${B^2 }$} \!\mathord{\left/
 {\vphantom {{{\bf B}^2 } {\beta ^2 }}}\right.\kern-\nulldelimiterspace}
\!\lower0.7ex\hbox{${\beta ^2 }$}}} } \right)} }}. \label{Expo60}
\end{equation}
Following the same steps that led to (\ref{Exp145}) we find that the $\left\langle H \right\rangle _\Phi ^{\left( 1 \right)}$-term is given by
\begin{equation}
\left\langle H \right\rangle _\Phi ^{\left( 1 \right)}  = \left\langle \Phi  \right|\int {d^3 x} \left\{ {\frac{1}{2}{\bf \Pi} ^2  - \frac{3}{{8\beta ^2 }}{\bf \Pi} ^4 } \right\}\left| \Phi  \right\rangle. \label{Expo70} 
\end{equation}
From Eqs. (\ref{Exp145}) and (\ref{Expo70}), it is clear that both expressions, up to a numerical factor, give rise to the same interaction energy.

\section{Final Remarks}

In summary, within the gauge-invariant but path-dependent variables formalism, we have considered the confinement versus screening issue for both Born-Infeld-like electrodynamics and exponential  electrodynamics. Once again, a correct identification of physical degrees of freedom has been fundamental for understanding the physics hidden in gauge theories. We should highlight the identical behaviors of the potentials associated to each of the models. Interestingly enough, their non-commutative version displays an ultraviolet finite static potential. The above analysis reveals the key role played by the new quantum of length in our analysis. In a general perspective, the benefit of considering the present approach is to provide unifications among different models, as well as exploiting the equivalence in explicit calculations, as we have illustrated in the course of this work.

More recently, an up-dated upper bound for the electron's electric dipole moment (EDM) has been published in \cite{nphoton}. Since the understanding of this property involves CP-violation, we believe it would be a viable task to include a CP-violating term given by $\cal G$, or an odd power of $\cal G$, and to compute how it may yield an asymmetric charge distribution around the spin of the electron. This, in turn, should induce a contribution to the electron's EDM in the framework of the Born-Infeld model. To do that, it is clearly important to also know the magnetic field that appears as an effect of the non-linearity in the case of a point charge \cite{DeAssis:2010cv}. We are presently pursuing this investigation we hope to report on it soon.

\begin{acknowledgments}
P. G. was partially supported by Fondecyt (Chile) grant 1130426, DGIP (UTFSM) internal project USM 111458. P. G. also wishes to thank the Field Theory Group of the CBPF for hospitality and PCI/MCT for support.
\end{acknowledgments}

\end{document}